\documentclass[twocolumn]{article}
\usepackage{chase}

\usepackage[utf8]{inputenc} 
\usepackage[T1]{fontenc}    
\usepackage{hyperref}       
\usepackage{url}            
\usepackage{booktabs}       
\usepackage{amsfonts}       
\usepackage{nicefrac}       
\usepackage{microtype}      
\usepackage{lipsum}
\usepackage{graphicx}
\usepackage{amsmath}
\usepackage{amssymb}

\usepackage{xcolor}
\usepackage{relsize}
\usepackage{caption}
\usepackage{subcaption}

\usepackage{appendix}

\usepackage[capitalize]{cleveref}

\newcommand{\pt}{p_\mathrm{T}}

\title{Particle Convolution for High Energy Physics}

\author{
  Chase Shimmin \\
  Department of Physics\\
  Yale University\\
  New Haven, CT 06511 \\
  \texttt{chase.shimmin@yale.edu} \\
}

\begin{document}
\maketitle

\begin{abstract}
We introduce the Particle Convolution Network (PCN), a new type of equivariant neural network layer suitable for many tasks in jet physics.
The particle convolution layer can be viewed as an extension of Deep Sets and Energy Flow network architectures, in which the permutation-invariant operator is promoted to a group convolution.
While the PCN can be implemented for various kinds of symmetries, we consider the specific case of rotation about the jet axis the $\eta - \phi$ plane.
In two standard benchmark tasks, q/g tagging and top tagging, we show that the rotational PCN (rPCN) achieves performance comparable to graph networks such as ParticleNet.
Moreover, we show that it is possible to implement an IRC-safe rPCN, which significantly outperforms existing IRC-safe tagging methods on both tasks.
We speculate that by generalizing the PCN to include additional convolutional symmetries relevant to jet physics, it may outperform the current state-of-the-art set by graph networks, while offering a new degree of control over physically-motivated inductive biases.
\end{abstract}

\begin{figure*}[h!]
    \centering
    \includegraphics[width=\textwidth]{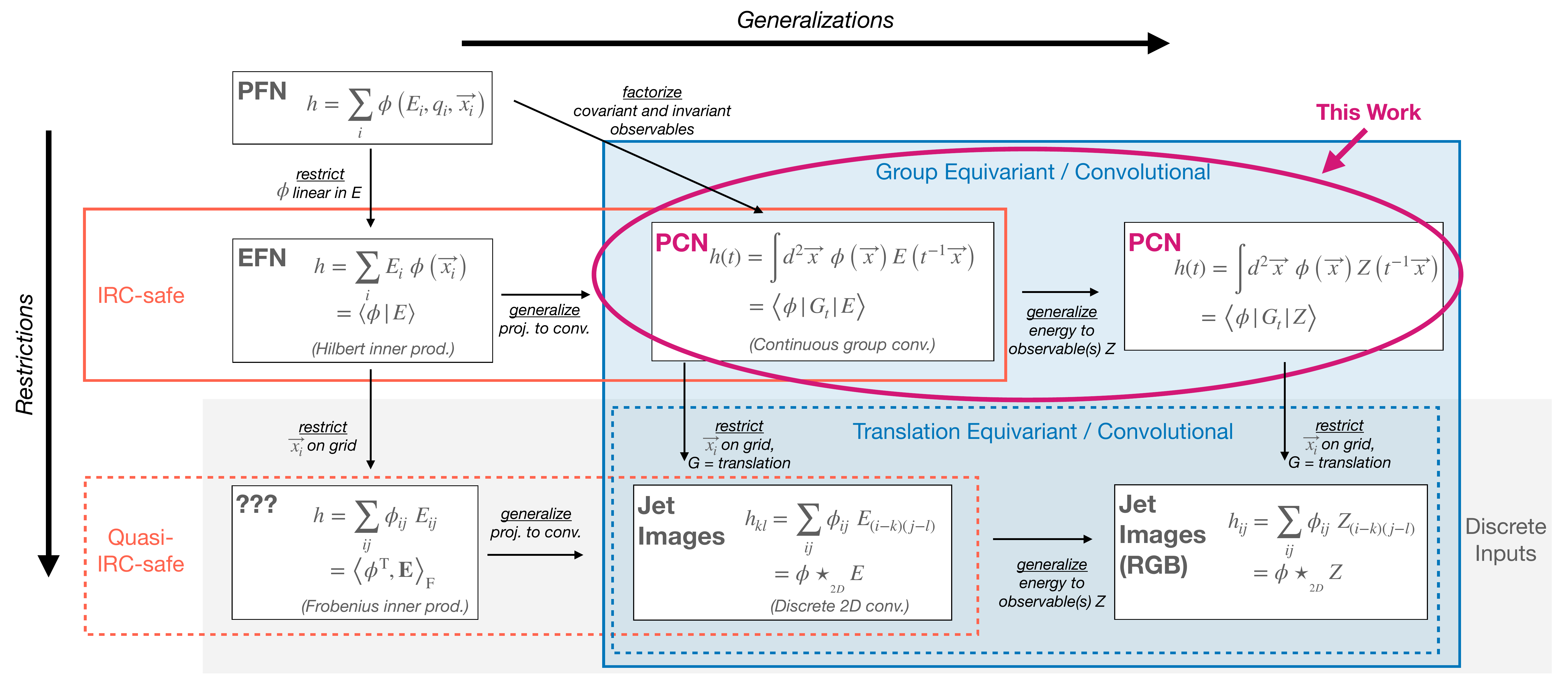} 
    \caption{
    Conceptual map indicating the relationship between ParticleFlow, EnergyFlow, and Jet Image architectures. Particle Convolution (this work) can be seen as a generalization of jet images to continuous space. It can also be seen as a generalization of the EnergyFlow network, where Hilbert space projection has been promoted to (arbitrary) group convolution.
    Mathematical definitions are provided in section~\ref{sec:particle-projection}.
    }
    \label{fig:jetland}
\end{figure*}

\section{Introduction}
A common problem in high energy physics is the desire to use state-of-the art machine learning methods to solve real-world problems encountered in physics experiments.
Often, the leading edge of machine learning is driven by a standard suite of problems which are motivated by the practical interests of tech industries.
For example, convolutional neural networks (CNNs) for image processing and Recurrent Neural Networks (RNNs) for natural language processing are some of the most well-studied and developed areas in deep learning.
In an effort to reap the benefits of these advances, physicists have been systematically studying the efficacy of these existing methods when applied to typical problems in our field such triggering, event reconstruction, and object tagging.\cite{Cheng:2017rdo,Egan:2017ojy,Luo:2017ncs,deOliveira:2015xxd,Cogan:2014oua,Guest:2018yhq,CMS-DP-2017-049,Shlomi2020GraphNN}

However, the structure and underlying processes of the data for which these architectures were developed are often fundamentally different from those present in physics experiments.
In order to bridge this gap, physicists have often resorted (with a few exceptions\cite{Louppe:2017ipp,pmlr-v119-bogatskiy20a,Moreno:2019bmu,Komiske:2018cqr}) to remedial measures by re-structuring or re-formatting experimental data from its natural representation to conform with the format required by a specific existing model architecture.

We consider the problem of jet-tagging\cite{Asquith:2018igt,Gallicchio:2011xq,Larkoski:2017jix}, in which the goal is to discriminate between various types of hadronic decays based on the detailed structure of energy and tracking measurements from the detector.
Hadronic jets are the most abundant products of proton collisions at the LHC and are formed when energetic quarks or gluons fragment recursively into lower-energy quarks and gluons until stable particles form, a process known as hadronization.
These particles are then observed by the particle detector.
Experimentally, jets are defined by a specific clustering algorithm\cite{Cacciari:2008gp} which groups together energy deposits localized within a certain angular scale set by a chosen radius parameter.
Jets can be initiated directly from hard QCD processes as well as rare particle decays\cite{Plehn:2009rk,Abdesselam:2010pt,Butterworth:2008iy,Shimmin:2016vlc}.
The exceedingly large QCD cross sections often lead to a scenario where the rate of background jets overwhelms the signal of interest.

In this work, we will consider two common tasks: quark/gluon (q/g) identification\cite{Komiske:2018vkc,Gallicchio:2011xq,Dasgupta:2007wa,Gallicchio:2012ez,Larkoski:2014pca}, and top-quark tagging\cite{Banfi:2006hf,Larkoski:2017jix,Asquith:2018igt}.
In q/g identification, the goal is to determine whether a jet originated from a final-state quark or gluon, which subsequently fragmented into the pattern of radiation observed by the detector.
Top quarks decay before hadronization can occur, and frequently result in three collinear jets, which may overlap significantly in the detector when produced at high momentum.
The goal of top-tagging is to determine whether or not a large-radius jet pattern originated from a top quark.

In data, a jet is most naturally represented by a variable-sized collection of four-vectors representing momentum as measured by calorimeter deposits and/or charged particle trajectories.
These vectors may also be annotated with additional information, such as charge or particle type.
The precision with which these constituents are measured depends on the detector properties.
For example, charged particle trajectories yield very precise directional measurements but do not include neutral particles, while calorimeter deposits have coarser directional precision but better energy precision.

While the field of HEP has used certain machine learning methods, such as Multilevel Perceptrons\cite{Kolanoski:1995zn,Bass:1996ez,Milek:1999xm} and Boosted Decision Trees\cite{Roe:2004na}, for decades, these models are not directly amenable to the data in question.
Instead, physicists have constructed a large number of mathematical and heuristic \textit{jet substructure observables} which quantify various properties on a per-jet level.\cite{Thaler:2010tr,Larkoski:2014pca,Komiske:2017aww,Moult:2016cvt,Larkoski:2014gra,Dasgupta:2013ihk,Larkoski:2014wba}
These observables may then be used in standard multivariate analysis techniques.

Developments in the field of deep learning have led to a reexamination of the problem of jet tagging, as new architectures have emerged which can operate on constituent-level data.
Alternative proposals include casting the particle data into a 2D pixel image in order to apply to a CNNs\cite{deOliveira:2015xxd,Cogan:2014oua}, passing the particle's features one at a time as a sequence into an RNN\cite{Egan:2017ojy,Cheng:2017rdo,Louppe:2017ipp}, or embedding groups of nearby constituents into a graph for use in a Graph Neural Networks\cite{Qu:2019gqs}.
While RNNs in particular have seen successful application in experiments, more recently the field has been moving to permutation-invariant set based networks\cite{2017arXiv170306114Z,Komiske:2018cqr} which are easier to train and achieve comparable performance.

Until now, the question of equivariance properties has not received much attention from the field.
The CNN-based jet image approach, which respects an approximate 2D translational symmetry, was for a long time the only example of an equivariant architecture that had been studied.
While translation equivariance has proved highly useful in image processing tasks, it is unclear whether it is physically meaningful in the context of jet substructure.
Very recently, it has been proposed to consider a more physically-relevant class of equivariance, specifically, the Lorentz Group Network\cite{pmlr-v119-bogatskiy20a}. The LGN architecture operates at the constituent level and is fully equivariant with respect to arbitrary lorentz transformations.
However, the LGN has not yet proved to work as well as existing methods, possibly due to the exceedingly large memory structures required to implement sufficiently complex networks.

In this paper, we turn our attention to the curious gap between the PFN and CNNs.
PFNs, which are the current state-of-the-art in most experimental applications, generally perform as well as any other methods (with the possible exception of the newly-proposed GNNs), but they possess no particular type of equivariance.
On the other hand, the CNN approach which does exhibit equivariance, is generally outperformed by the other methods mentioned, possibly due to the sparsity/discretization or due to the equivariance being of the wrong type.

It turns out there is a specific connection between the PFN and the CNN architectures.
They can be viewed from a common mathematical perspective, illustrated schematically in Fig.~\ref{fig:jetland}.
This mathematical connection, which is elaborated in Sec.~\ref{sec:particle-projection} and \ref{sec:rotational-convolution}, is essentially that with a subtle modification, PFNs can be viewed as the geometrical operation of projection.
This projection operation then can be easily promoted to a convolution, which we call Particle Convolution.
The Particle Convolution allows us to build networks which feature equivariance with respect to a much larger class of symmetry groups, while operating directly on the constituent-level data.
We show that the jet image method is a special case of Particle Convolution in the case of a discrete shift operator and binned coordinates, while the EFN is a special case in which the operator is nullary.

In Sec.~\ref{sec:equivariance} we will discuss the concept of equivariance, and motivate the particular case of rotational equivariance for the problem of jet tagging.
We then consider in Sec.~\ref{sec:particle-projection} a formal connection between the permutation-invariant set-based models and the notion of Hilbert space projection.
In Sec.~\ref{sec:rotational-convolution}, we demonstrate how to promote these projective operations to convolutions possessing equivariance properties by construction, using the particular example of rotation.
In Sec.~\ref{sec:steerable-convolutions}, we consider some of the technical challenges in implementing Particle Convolution, and detail a more efficient solution based on the notion of steerable functions.
In Sec.~\ref{sec:experiments} we provide details for experiments conducted on the two benchmark tasks, q/g tagging and top-tagging, and in Sec.~\ref{sec:conclusion} we present results and conclusions.

\section{Equivariance}
\label{sec:equivariance}

In this section, we begin with a formal definition of the mathematical concept of equivariance.
Then, before proceeding to the technical details of how this definition can be applied, we present an intuitive argument for how equivariance can benefit machine learning models in the specific application of jet tagging.

A map $f : X\rightarrow Y$ is \textit{equivariant} with respect to a group $G$ acting on $X$ if for every $g\in G$, there exists some $\Pi_g : Y \rightarrow Y$ such that:
\begin{equation}
    f(g \cdot x) = \Pi_g f(x)\,,\ \forall x \in X \,.
\end{equation}
In other words, given an equivariant function $f$, it is possible to determine the result of $f(g \cdot x)$ by applying either $g$ to the input, or $\Pi_g$ to the output of the function.
Note that invariance is a special case of equivariance: the function $f$ is said to be invariant when $\Pi_g$ is equal to the identity for every $g$.

The most commonly known equivariant neural architecture is the Convolutional Neural Network (CNN).
These networks are equivariant with respect to discrete translational shifts in one or more dimensions.
In particular, two dimensional CNNs excel at computer vision tasks.
The intuitive reason behind this is that CNNs can learn generic features, such as textures or edges, and is able to match any of its learned features at any location on an image.
If a particular feature is shifted to another location in the image, the CNN's response to that feature will be shifted a corresponding amount.
In contrast, a simple fully-connected neural network would need to re-learn a new instance of each feature

Many recent developments in the field of machine learning have focused on the analysis and design of equivariance properties of neural networks.
In many cases, equivariance can be generalized to aribrary homogeneous spaces via group convolution, and these architectures often lead to improved performance when a relevant symmetry of the data can be exploited\cite{cohen2019general,Weiler2018LearningSF,2018arXiv180208219T}.

In the context of jet tagging, we propose to consider the specific case of rotation.
The \textit{rotational} Particle Convolution Network (rPCN), detailed in Sec.~\ref{sec:rotational-convolution}, evaluates a different kind of convolution with respect to rotation about the jet axis in the $\eta - \phi$ plane.
This case is physically relevant and also happens to be mathematically simple.
The rationale for this particular equivariance stems from the physical processes governing jet formation, which are approximately invariant under rotation about the jet axis.
Therefore, common features may emerge in radiation patterns which differ only in arbitrary rotation about the jet axis.

To address this, some works \cite{deOliveira:2015xxd} have proposed removing the rotational degree of freedom via a pre-processing step that imposes a standardized reference orientation for all jets.
This technique can in some cases improve performance of non-equivariant models, although in other cases it can have a detrimental effect.
In any case, it effectively manages to render the entire model invariant under \textit{global} rotations.

In many applications we do in fact desire an invariant network response; for example, when determining whether a jet originated from a boosted $Z$ boson decay, we expect the same answer regardless of the random orientation of the parent particle's decay axis.
However, rather than constructing an invariant input to eliminate the rotational variation in the data, we can instead attempt to preserve the structure of the underlying symmetry within the network itself by enforcing equivariance at each layer.
This allows a deep network to build rich representations to examine and compare the angular structure of various features.

In the following sections, we will describe one way in which a rotationally-equivariant network can be constructed.
This architecture, which we call a Particle Convolution Network, can be generalized to effect equivariance with respect to additional types symmetry as well.

\begin{figure}[ht]
    \centering
    \includegraphics[width=0.5\textwidth]{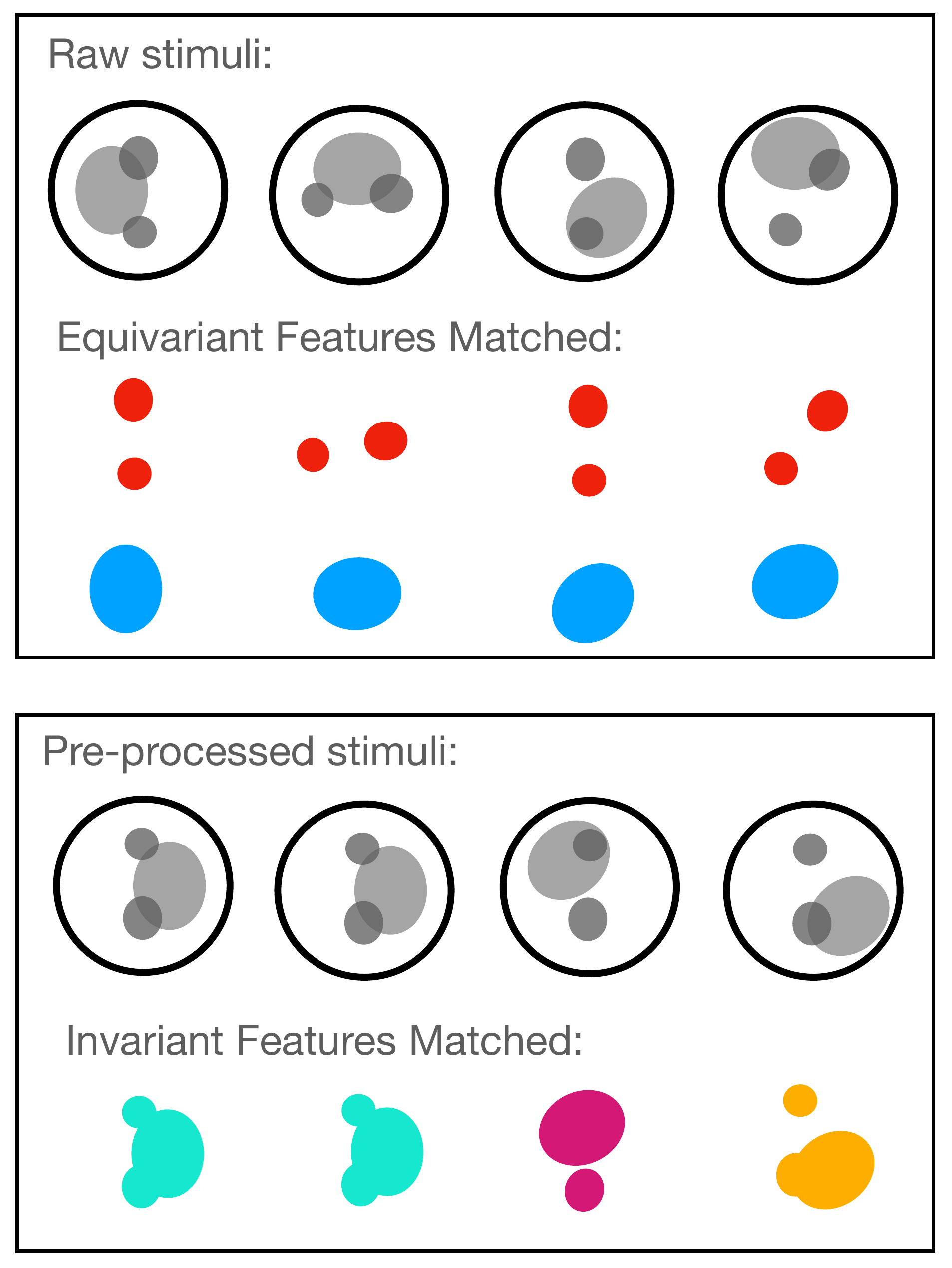}
    \caption{Illustration of invariant vs. equivariant feature matching.
    Colors correspond different learned filter channels.
    In this example, the input stimuli are a superposition of two radiation patterns, each of which has an arbitrary angular rotation.
    The invariant approach is able to match inputs differing by a global rotation; however, it must learn different features for each relative orientation.
    A network with equivariant filters can match each pattern regardless of relative orientation.
    }
    \label{fig:rotation}
\end{figure}

Figure~\ref{fig:rotation} illustrates this point.
In this example, we consider two independent features that might appear in a jet: a diffuse blob of low-energy particles, and a pair of high-$\pt$ subjets.
If the diffuse particles and the hard subjets are produced at different stages during parton showering, their relative angular orientations about the jet axis could be largely random and uncorrelated.
Rotational pre-processing does effectively limit the amount the network needs to learn in the case where two jets differ only by a global rotation.
However, such a network still needs to learn a completely different set of features to recognize various relative orientations between the hard and soft particles.
On the other hand, an equivariant network might learn a single feature representing ``two subjets'' and one feature for ``diffuse lobe'', and by construction understands that these two features independently have a rotational degree of freedom.
Having learned such features, the equivariant model would easily detect either element of the substructure at arbitrary angles, and can pass this information to deeper layers within the network.
Subsequent layers could then proceed to execute computations which reason about either the absolute or relative angular position of the features.

This is precisely the intuition behind the rotational convolution: features in the jet may be matched by evaluating the projection of the jet onto a series of filters representing learned features.
These projections are sampled along a range of rotational orientations, resulting in a periodic ``waveform'' representing each individual filter's response as a function of angle.
At this point, the network has formed a representation consisting of a discrete 1D waveform with multiple channels corresponding to the different filters.

\section{Particle Projection}
\label{sec:particle-projection}

In order to define the Particle Convolution, we begin by examining the structure of the ``Deep Sets''-based Energy Flow and Particle Flow networks (EFN and PFN).
These networks operate on a jet $S$ represented by a set $S = \{s_1, \dots, s_{|S|}\}$ of observations $s_i$.
We shall assume the observations $s_i = (\vec{x}_i, q_i)$ are composed of a 2D coordinate $\vec{x}_i$ representing the direction of particles relative to the jet axis in the $\eta - \phi$ plane\footnote{To simplify notation, we will always assume jets have been centered in the $\eta - \phi$ plane, so that the coordinate $\vec{x} = (\Delta \eta, \Delta \phi) = (\eta, \phi)$.}, and some non-coordinate quantities $q_i$.
The non-coordinate observable $q_i=(e_i, \dots)$ is composed of at least an energy $e_i$ and may be supplemented by additional observables such as charge, mass, etc.

The EFN depends only on the particle coordinate $\vec{x}_i$ and energy $e_i$, and is defined as:
\begin{equation}
\label{eq:efn-def}
    \mathrm{EFN}(S) = F\left( \sum_{k=1}^{|S|} e_k \Phi(\vec{x}_k) \right) \,,
\end{equation}
where $F$ and $\Phi$ are arbitrary continuous functions, generally represented by neural networks.

Since the learnable function $\Phi$ is defined over (approximately) Euclidean spatial coordinates, we might consider it as an element of the Hilbert space $L^2(\mathbb{R}^2)$.
If we likewise consider the values of the non-coordinate observables of the jet $S$ to be represented as a function on $\mathbb{R}^2$, we can ``expand'' the jet in the continuous basis $\left| \vec{x} \right>$.
That is, for each collection $S$, we define an associated representation $\left| Z \right>_S$ such that
\begin{equation}
    \left<\vec{x} | Z \right>_S = \mathcal{Z}_S(\vec{x}) := \sum_{k=1}^{|S|} Z(q_k) \delta(\vec{x} - \vec{x}_k) \,,
\end{equation}
where $Z$ denotes some arbitrary function of the non-coordinate observable(s) $q$.
That is, in the coordinate basis, $\left| Z \right>_S$ corresponds to a function $\mathcal{Z}_S : \mathbb{R}^2 \rightarrow \mathbb{R}$ which represents the spatial distribution of the quantity $Z(q)$ within the jet $S$.
This function, $\mathcal{Z}_S$, is of course not continuous, and also highly sparse.
Strictly speaking, it is not an element of $L^2(\mathbb{R}^2)$, but rather a generalized function.
Nonetheless, we shall show that this analytic expression of the jet and its observables provides a useful mathematical tool.

For instance, if we consider an arbitrary continuous function $\Phi : \mathbb{R}^2 \rightarrow \mathbb{R}$, we may exploit the definition of the delta function in order to evaluate the Hilbert space inner product
\begin{align}
\label{eq:proj-z}
    \left< \Phi, Z\right>_S &= \int d^2\vec{x}\ \Phi(\vec{x}) \mathcal{Z}_S(\vec{x}) \\
    &= \int d^2\vec{x}\ \Phi(\vec{x}) \sum_k Z(q_k) \delta(\vec{x}-\vec{x}_k) \\
    &= \sum_k Z(q_k) \Phi(\vec{x}_k) \,.
\end{align}

In the special case $Z = \mathrm{E}$ where $\mathrm{E}(q_k) = e_k$ simply returns the energy of the particle $k$, we have:
\begin{equation}
\label{eq:proj-efn}
    \left< \Phi, \mathrm{E} \right>_S = \sum_k e_k \Phi(\vec{x}_k) \,.
\end{equation}
This last equality can immediately be recognized as the inner term of Eq.~\ref{eq:efn-def}.
In other words, we may consider the EFN to be an arbitrary function $F$ acting on the \textit{projection} of a jet's empirical energy distribution $\left| \mathrm{E} \right>_S$ onto a learned filter $\left| \Phi \right>$:
\begin{equation}
    \mathrm{EFN}(S) = F\big( \left< \Phi, \mathrm{E} \right>_S \big) \,.
\end{equation}

In the following section, we will exploit this interpretation in order to define general equivariant convolution operators.
But first, we will consider the PFN from this perspective.
The PFN as defined in~\cite{Komiske:2018cqr} is given by
\begin{equation}
    \mathrm{PFN}(S) = F\left( \sum_k \Phi(\vec{x}_k, q_k) \right) \,.
\end{equation}
Comparing with Eq.~\ref{eq:efn-def}, it is clear that the EFN is a special case of this PFN, where $\Phi$ is linear in the particle energy $e_k$.
It is this linearity which guarantees IRC-safety in the EFN case.
However, it is the fact that $\Phi$ depends only on $\vec{x}$ that allows the projection Eq.~\ref{eq:proj-z} to be related to the EFN.
This is because the inner product is defined in terms of an integral with respect to a meaningful topological measure -- in this case, two dimensional Euclidean space.
As we shall see, this is important for defining convolution with respect to locally compact groups, such as rotation and translation.

Therefore, in order to extend the concept of projection to the more general case, we define a modified version of the PFN directly in terms of the projection operator of Eq.~\ref{eq:proj-z}:
\begin{align}
    \mathrm{PFN}'(S) &= F\left( \left< \Phi, Z \right>_S \right) \\
    &= F\left( \sum_k Z(q_k) \Phi(\vec{x}_k) \right) \,.
\end{align}
Here, $F$, $Z$, and $\Phi$ are all arbitrary continuous functions that could be implemented via neural networks.
This network is equivalent to a PFN where the per-particle function is required to be separable into terms depending on the coordinate and non-coordinate observables.
In principle, this represents a strictly less general model than the original PFN.
However, in this form the $\mathrm{PFN}'$ readily admits generalization via convolution, which we shall find can result in much more effective models.

\subsection{Equivariant Projections}
\label{subsec:equivariant-projections}
Before proceeding to define the Particle Convolution, we first make some observations about the potential for equivariance in the EFN and $\mathrm{PFN}'$ architectures, which are based on projection.
In particular, it is straightforward to show that whenever $\Phi$ possesses a specific form of equivariance, so does its projection.
Suppose $\Phi$ is equivariant with respect to $G$ so that $\Phi(g\cdot x) = \Pi_g \Phi(x)$ for all $x \in X$ and $g \in G$. Then, if $\left|Z\right>_S \rightarrow g\left|Z\right>_S$, we have:
\begin{align}
    \left<\Phi | Z \right>_S \rightarrow &\left<\Phi | g | Z\right>_S \\
     = &\int dx\ \Phi(g \cdot x) \mathcal{Z}_S(x) \\
     = &\int dx\ \Pi_g \Phi(x) \mathcal{Z}_S(x) \\
     = &\ \Pi_g \left< \Phi | Z \right>_S \,.
\end{align}
So we can see that the projection is also equivariant to transformations $g$ applied to the particles of $\left|Z\right>_S$.

In practice this is often of limited utility.
For example, if $\Phi$ is constant on the $\eta - \phi$ plane, it is invariant \textit{w.r.t.} translations, and clearly so is the projection.
If $\Phi$ has some specific periodic behavior under rotations, so that in polar coordinates $\Phi(r, \theta + \delta) = \Phi(r, \theta)$ for some $\delta$, then the projection will also have this periodicity.
However, such filters are able to express only limited pattern-matching ability, and may not lead to sufficiently complex representations for the task at hand.

In the following section, we will see that by working with convolutions, arbitrary filters can be learned while retaining equivariance.

\section{Particle Convolution}
\label{sec:rotational-convolution}

Having re-cast the core operation of Particle Flow Networks in terms of the geometric concept of projection, we can immediately generalize the network by promoting projection to convolution.
In this section, we demonstrate this concretely with the example of rotational convolution.

The rotational Particle Convolution between a jet $\left|Z\right>_S$ and filter $\left<\Phi\right|$ is defined as the function:
\begin{align}
    h(\Delta;\ \Phi, Z, S) :=&\ [\Phi \star Z ]_S(\Delta) \\
    =& \left< \Phi | R_\Delta | Z \right>_S \\
    =& \int d^2\vec{x}\ \Phi(R_\Delta \vec{x}) \mathcal{Z}_S(\vec{x}) \\
    =& \sum_k Z(q_k) \Phi(R_\Delta \vec{x}_k) \,,
\end{align}
where $\Delta$ is the angle about the jet axis, and $R_\Delta \in SO(2)$ is the corresponding rotation operator.
We give the convolution the handle $h(\Delta)$, omitting the independent arguments $(\Phi,Z,S)$ when convenient, to emphasize that the result of the convolution is a \textit{function} of angle.

\begin{figure*}
    \centering
    \includegraphics[width=\textwidth]{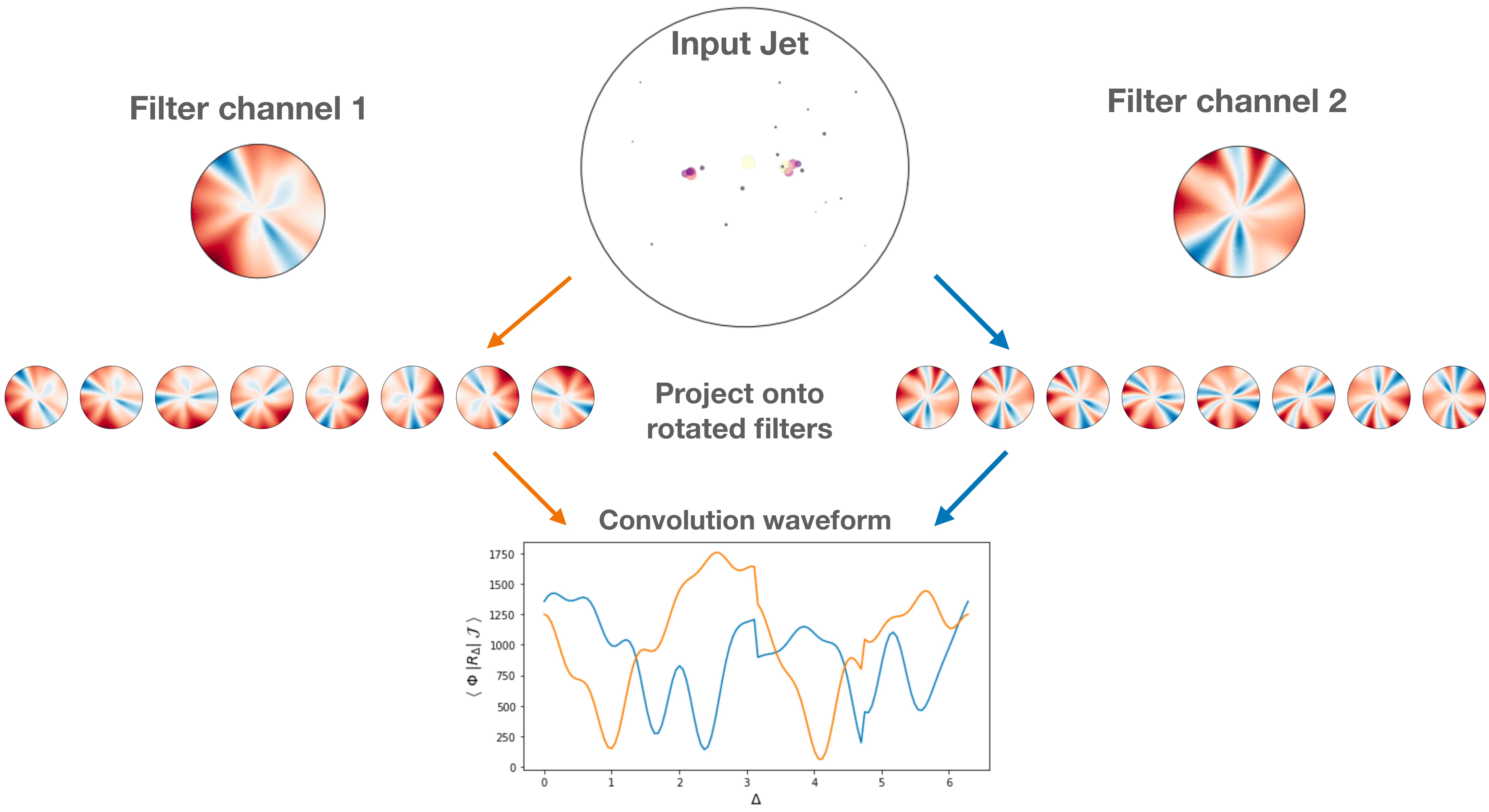}
    \caption{Illustration of the particle convolution of an example jet, represented as a point cloud, and two different filters.}
    \label{fig:pconv}
\end{figure*}

Let us examine the equivariance of this operation with respect to a rotation of the input particles $\left| Z \right>_S$:
\begin{align}
    |Z\rangle_S \rightarrow &R_\delta | Z \rangle_S \implies \\
    h(\Delta) \rightarrow &\langle \Phi | R_\Delta \big[\ R_\delta | Z\rangle_S \big] \\
    =&\langle \Phi | R_{\Delta + \delta} | Z \rangle_S \\
    =&h(\Delta + \delta) \\
    =&T_{-\delta}\ h(\Delta) \,,
\end{align}
where $T$ is the coordinate shift operator acting on the function $h$, defined by $T_y f(x) = f(x-y)$.

Since this is true for every $R_\delta \in SO(2)$ and every $\Delta$, we have established the equivariance of the convolution $h(\Delta)$ with respect to the group $SO(2)$ acting on the jet $\left| Z \right>_S$.
Moreover, we see that rotation of the input particles corresponds to a shift of the output convolution.

This convolution operation comprises the first layer of the rPCN.
The network parameters of this layer are encoded by the functions $\Phi$ and $Z$, which could for instance be themselves neural networks.
Generally, we would like to pass the result of this first layer to additional layers in a deep neural network.
However, it is not clear what to do with a continuous function.

In practice, the convolution can be \textit{sampled} at $n$ discrete points $\Delta_i = 2\pi i/n$.
In this case, we can represent the convolution $h$ as a tensor with index $i$ representing the sampled points of the convolutional ``waveform'':
\begin{equation}
\label{eq:conv-sample}
    h_i = \left<\Phi | R_{\Delta_i} | Z \right>_S = \sum_k Z_S(q_k) \Phi(R_{\Delta_i} \vec{x}_i) \,.
\end{equation}
The sampled convolution is now, strictly speaking, equivariant with respect to the subgroup of discrete rotations by $2\pi i/n$:
\begin{align}
    \left|Z\right>_S \rightarrow &R_{\frac{2\pi j}{n}} \left| Z \right>_S \implies \\
    h_i \rightarrow &\left< \Phi \right| R_{\frac{2 \pi (i+j)}{n}} \left| Z \right>_S = h_{i+j} \,.
\end{align}
Hence, a discrete rotation of the particles in $S$ corresponds to a cyclic permutation of the indices $h_{i+j}$.

By sampling a larger number of points $n$, the network can approximate continuous equivariance.
Conversely, by limiting the number of samples in accordance with the Shannon-Nyquist theorem, the network can be designed so as to ignore high frequency information which might be considered noise.

Having obtained a shift-equivariant tensor $h_i$, it is straightforward to build deeper equivariant representations by apply the standard 1D CNN layer.
The discrete 1D convolution layer is shift equivariant; however, care must be taken to also enforce the cyclic boundary conditions.
This can be done by appropriate padding of the inputs: for a convolution with kernel size $k$, the tensor $h = (h_1, \dots, h_n)$ should be extended as follows:
\begin{equation}
    h^+ = ( h_{(n-k+1)/2}, \dots, h_{n} \Big| h_1, \dots, h_n \Big| h_1, \dots, h_{(k-1)/2} ) \,.
\end{equation}
Note that in contrast to conventional CNNs, the number of samples in the convolution output is not reduced, but rather stays the same due to the periodic boundary condition.
However, it is possible to reduce the tensor along its sample axis via downsampling.

After processing by any number of additional convolutional layers, an invariant representation may be formed by a global pooling operation.
For example, it is clear that taking the maximum or average value of $h_i$ along the sample axis yields an invariant quantity.
After the pooling operation, each filter channel yields a single quantity describing an invariant feature of the input jet.
Once the invariant is formed, any additional functions applied will also result in invariants.
For example, the collection of invariant filter responses may be passed to densely connected layers, and finally to whatever output is suitable for the task objective.

\section {Steerable Convolutions}
\label{sec:steerable-convolutions}

In Sec.~\ref{sec:rotational-convolution}, we defined the rotational Particle Convolution operation.
This convolution can in principle be sampled by applying a series of rotations to the particle coordinates $R_{\Delta_i}\vec{x}_k$, and re-evaluating $\Phi$ in the projection of Eq.~\ref{eq:conv-sample}.
In practice, this can be problematic as a sizeable neural network may be required to model $\Phi$.
For example, in some experiments described in Sec.~\ref{sec:experiments}, we consider architectures in which $\Phi$ must be applied to up to 150 particles, and re-sampled at up to 21 orientations.
Therefore, even at a moderate batch size of 64 jets, a single network layer of 128 units results in a 32-bit tensor occupying nearly one GiB in memory.
This means that models based on direct sampling of convolutions cannot scale well due to memory limitations of current GPUs, and are also very time-intensive to train and optimize.

In this section, we show how the convolution may be implemented more efficiently using \textit{steerable functions}\cite{cheng1998theory,Weiler2018LearningSF}.
A function is said to be steerable if it can be expressed as a linear combination of equivariant functions\cite{cheng1998theory}.
By structuring a PCN such that the learnable filters $\Phi$ are expressed in an appropriate equivariant basis, it is possible to efficiently sample convolutions at arbitrary points by evaluating $\Phi$ only once.

This works by imposing some particular structure on our learnable filters.
Again, we will demonstrate with the specific case of rotation.
Consider for example, a filter $\psi$ of the form:
\begin{equation}
\label{eq:phi-m-def}
    \phi_m(\vec{x}) = \rho_m(r) e^{i m \theta} \,,
\end{equation}
where $m$ is an arbitrary integer, $\rho_m$ is a arbitrary function, and $r$ and $\theta$ are polar coordinates about the jet axis in the $\eta - \phi$ plane.
For any filter of this form, we can see that the rotation operator acts as:
\begin{equation}
\label{eq:phi-m-equivariance}
    \phi_m(R^{-1}_{\Delta} \vec{x}) = \rho_m(r) e^{i m (\theta - \Delta)} = e^{-i m \Delta} \phi_m(\vec{x}) \,.
\end{equation}

We can easily sample the convolution with this function at any point $\Delta$ in terms of the un-rotated projection:
\begin{align}
    h(\Delta; \phi_m, Z, S) &= \left<\phi_m | R_{\Delta} | Z \right>_S \\
    &= \sum_k \phi_m(R_\Delta \vec{x}_k) Z(q_k) \\
    &= \sum_k e^{i m \Delta} \phi_m(\vec{x}_k) Z(q_k) \\
    &= e^{i m \Delta} \left<\phi_m | Z \right>_S \,.
\end{align}
In other words, we need only compute $h(0)$ once and then use $h(\Delta) = e^{im\Delta} h(0)$.
This is, of course, an example of an equivariant projection as described in Sec.~\ref{subsec:equivariant-projections}.

Unfortunately, the filter $\phi_m$ has a definite angular frequency $m$.
As mentioned in Sec.~\ref{subsec:equivariant-projections}, the pattern-matching ability of such a filter is quite limited.
However, this suggests we could exploit a similar type of equivariance by imposing a specific structure related to Eq.~\ref{eq:phi-m-def} to construct a more general $\Phi$.

\begin{figure*}[h!]
    \centering
    \includegraphics[width=\textwidth]{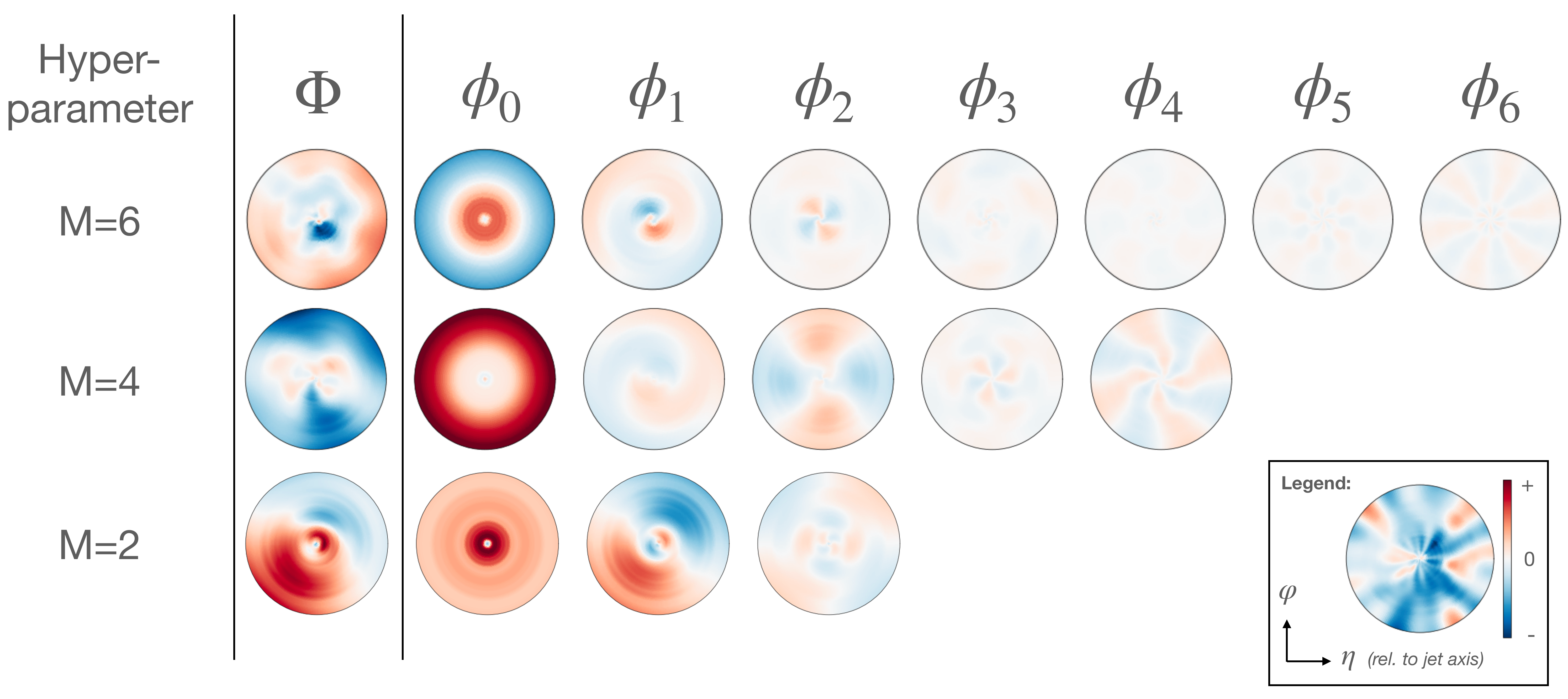} 
    \caption{
    Visualization of the series representation of learned filter instances $\Phi(\vec{x}) = \sum_m \phi_m(r, \theta)$ for various values of the cutoff hyperparameter $M$.
    }
    \label{fig:series}
\end{figure*}

In particular, let us re-define $\Phi$ in terms of a series of functions $\phi_m$:
\begin{equation}
\label{eq:phi-series}
    \Phi(\vec{x}) = \Phi(r, \theta) = \sum_{m=-M}^{M} \phi_m(r, \theta) = \sum_m \rho_m(r) e^{i m \theta} \,.
\end{equation}
Here, the hyperparameter $M$ represents a cutoff for angular frequencies captured by the filter.
The $2M+1$ radial functions $\rho_m$ are now the arbitrary learnable components of $\Phi$, which could be represented by ordinary neural networks, radial basis functions, \textit{etc}.
The relationship between $\Phi$ and this hyperparameter is depicted in Fig~\ref{fig:series}.

Note that in Eq.~\ref{eq:phi-series}, $\Phi$ is generally complex-valued, and so are the functions $\rho_m$.
For our purposes, we shall constrain it to be real-valued by imposing the condition $\rho_{-m} = \rho_{m}^\star$.
Alternatively, we could define $\Phi$ in terms of sines and cosines; however, the present form allows for simpler mathematical manipulation.

Each of the component functions $\phi_m$ are rotationally equivariant via Eq.~\ref{eq:phi-m-equivariance}.
However the filter $\Phi$, which combines arbitrarily many frequency modes, does not.
The trick is to instead associate $\Phi$ with a column vector
\begin{equation}
\Tilde{\Phi} = (\phi_{m}, ..., \phi_{-m})^\mathrm{T} \,,
\end{equation}
related via:
\begin{equation}
    \Phi(\vec{x}) = \mathbf{1}^\mathrm{T} \Tilde{\Phi}(\vec{x}) = \sum_m \phi_m(\vec{x}) \,.
\end{equation}
Now under a rotation $R_\Delta$, we have:
\begin{align}
    \Tilde{\Phi}(R^{-1}_\Delta \vec{x}) &= \big(\phi_M(R^{-1}_\Delta \vec{x}), \dots, \phi_{-M}(R^{-1}_\Delta \vec{x})\big)^\mathrm{T} \\
    &= \big( e^{-iM\Delta}\phi_M(\vec{x}), \dots, e^{iM\Delta}\phi_{-M}(\vec{x}) \big)^\mathrm{T} \\
    &= \mathbf{A}(\Delta) \Tilde{\Phi}(\vec{x}) \,,
\end{align}
where the square matrix
\begin{equation}
    \mathbf{A}(\Delta) = \mathrm{diag}\{e^{-iM\Delta}, \dots, e^{iM\Delta} \} \,.
\end{equation}
In other words, while the function $\Phi$ is not equivariant, the vector of functions $\Tilde{\Phi}$ \textit{is}, transforming as $\Tilde{\Phi} \rightarrow \mathbf{A}(\Delta)\Tilde{\Phi}$ under a rotation of coordinates by $\Delta$.

Therefore, in order to perform a convolution with $\Phi$, the network should compute the projection:
\begin{equation}
    \langle \Tilde{\Phi} | Z \rangle_S := \big(\left< \phi_M | Z\right>_S, \dots, \left< \phi_{-M} | Z\right>_S\big)^\mathrm{T} \,,
\end{equation}
once for the un-rotated case, and sample additional rotations at points $\Delta_i$ via
\begin{equation}
\label{eq:h-interpolation}
    \Tilde{h}_i = \langle\Tilde{\Phi} | R(\Delta_i) | Z \rangle_S = \mathbf{A}(\Delta_i) \langle\Tilde{\Phi}|Z\rangle_S \,.
\end{equation}
Finally, the sampled vectors $\Tilde{h}_i$ can be ``collapsed'' into a single numerical value by summing over the $m$-components:
\begin{equation}
    h_i = \mathbf{1}^\mathrm{T} \Tilde{h}_i
\end{equation}
Note that given a fixed set of sample points $\Delta_i$, the operators $\mathbf{A}(\Delta_i)$ are simply constant, diagonal matrices, and the matrix multiplication of Eq.~\ref{eq:h-interpolation} is a trivial operation for GPUs.

\begin{figure}[ht]
    \centering
    \includegraphics[width=0.5\textwidth]{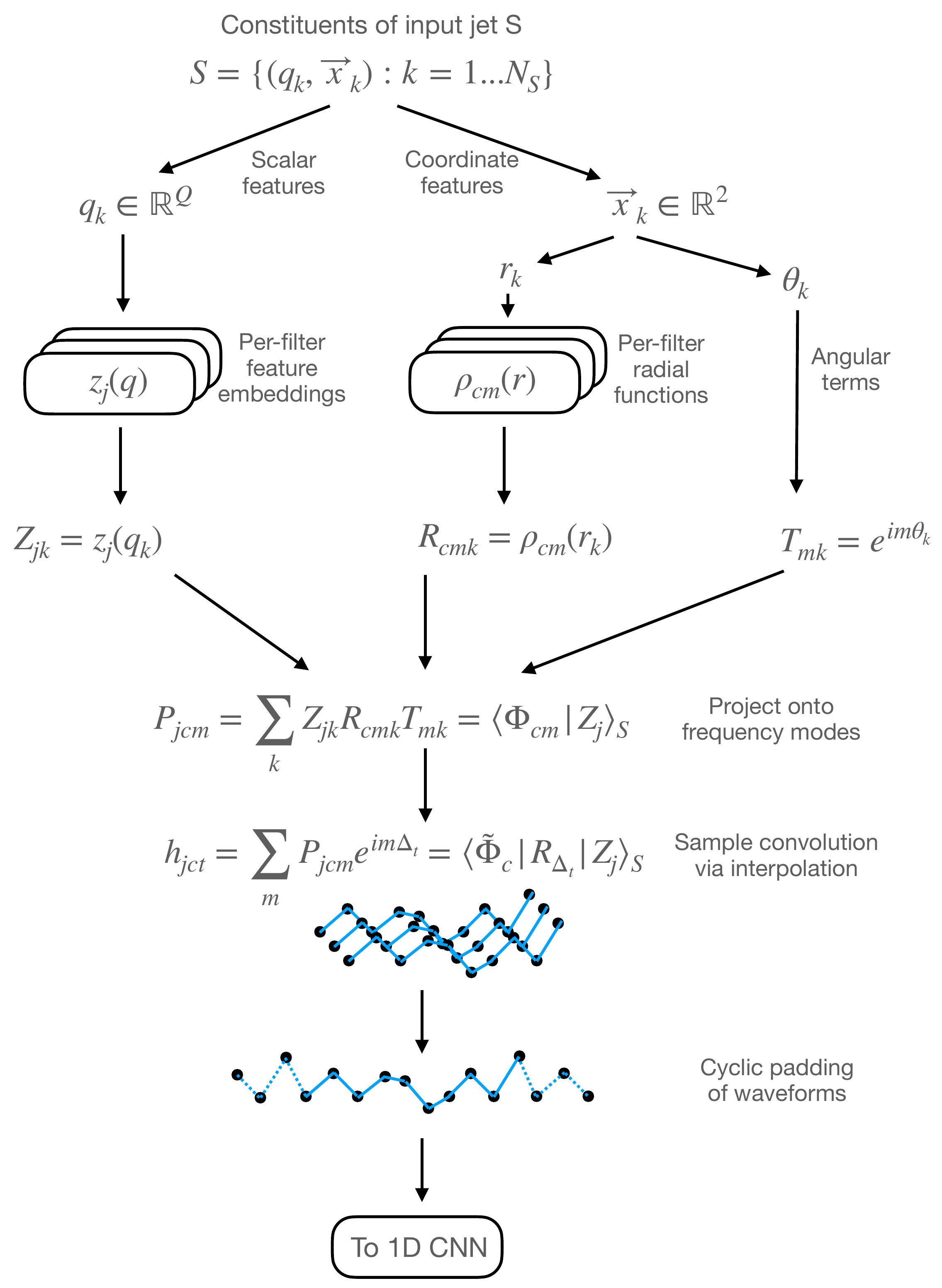}
    \caption{Schematic representation of the rotational Particle Convolution layer as described in Sec.~\ref{sec:steerable-convolution}.
    The functions $z_c$ and $\rho_{cm}$ can be specifically chosen (e.g. when setting $z_c = p_\mathrm{T}$ for IRC-safety), or can be implemented as neural networks.
    }
    \label{fig:flowchart}
\end{figure}

The network based on steerable convolutions is depicted schematically in Figure~\ref{fig:flowchart}.

Having re-formulated our filter $\Phi$ in terms of the equivariant functions $\phi_m$, we can revisit the question of efficiency and scalability for practical networks.
Note that in either case, for a filter $\Phi$ to capture information about given angular frequency $M$, we require $2M+1$ samples for $h_i$.
In Eq.~\ref{eq:conv-sample}, we do this by re-evaluating the projection with $\Phi$ $2M+1$ times.
In Eq.~\ref{eq:h-interpolation}, we evaluate the projection only once, but there are $2M+1$ functions $\phi_m$ which must be evaluated.
Therefore, it may seem that nothing has been gained.

However, in practice, the functions $\rho_m$ can be collectively implemented by a single neural network which shares most of its weights.
As long as the number of samples required, $2M+1$, is generally less than the size of hidden layers for typical dense networks, then the network should have a substantially smaller memory footprint.
This is because the intermediate tensors for the hidden layers need only be computed for the un-rotated case, as opposed to being calculated at every orientation as in Eq.~\ref{eq:conv-sample}.
This allows for smaller networks which are faster to train.

Moreover, the functions $\rho_m$ only need to learn a radial profile, rather than a response on the full 2-dimensional $\eta - \phi$ plane.
Therefore, it is reasonable to expect that simpler, smaller networks could be able to achieve equivalent results.
Lastly, since $\Phi$ in Eq.~\ref{eq:phi-series} is expressed in terms of functions with a definite cutoff frequency $M$, this acts as low-pass filter for angular structure.
Without this cutoff, higher frequency information may be aliased to lower modes during sampling, which could lead to a network sensitive to undesirable artifacts.

We find empirically that in either the direct sampling or steerable case, rPCNs of comparable angular resolution and depth achieve similar performance.
Therefore, when optimizing hyperparameters for the experiments in section~\ref{sec:experiments}, the more efficient steerable convolution architecture is used; however, it is possible that a more exhaustive optimization would find better results via the direct sampling approach.

\section{Experiments}
\label{sec:experiments}
We conduct experiments in training the rotational Particle Convolution Network for two benchmark problems in jet physics: quark/gluon identification, and top tagging.
The details of these datasets are described in Sec.~\ref{subsec:datasets}.
For both tasks, we train a general PCN as well as the IRC-safe variant.
The coordinates $\vec{x}$ are centered about the jet axis in the $\eta - \phi$ plane, as described in~\cite{Komiske:2018cqr}, and the scalar feature for each particle is the particle's transverse momentum, \textit{i.e.} $q_k = p_{\mathrm{T},k}$, which is invariant w.r.t. the coordinate-centering operation.
In the case of the q/g tagging task, we also test the performance of models when supplied with per-particle identification information as well, $q_k = (p_{\mathrm{T},k}, \mathrm{PID}_k)$.

\subsection{Datasets}
\label{subsec:datasets}

For reference, we consider two typical benchmark problems: q/g identification and top-quark tagging, using the same datasets as in Refs.~\cite{Komiske:2018cqr,Qu:2019gqs}.
The q/g dataset, described in Ref.\cite{Komiske:2018cqr} is comprised of $Z(\nu\nu) + q$ events (signal) and $Z(\nu\nu)+g$ events (background).
The events are simulated using \textsc{Pythia}8\cite{Sjostrand:2014zea}, which performs parton showering and hadronization.
No detector simulation is performed for this sample.
Particles (excluding neutrinos) are clustered using the anti-$k_\mathrm{T}$ algorithm\cite{Cacciari:2008gp} with radius parameter $R=0.4$.
The 4-momenta and particle ID of particles within the leading jet are saved for jets with $p_\mathrm{T}\in[500, 550]$ GeV and $|\eta|<2$.
The dataset is split into 1.2 million training events and 400k each of validation and test events.
In our experiments, we truncate events with greater than 68 particles by discarding those with the lowest $p_\mathrm{T}$.

For the top-tagging benchmark, we use the dataset available at Ref.\cite{kasieczka_gregor_2019_2603256}.
The events are generated with \textsc{Pythia}8 and passed into Delphes\cite{deFavereau:2013fsa} fast detector simulation, without pileup.
The signal process is $t\bar{t}$ and the background is inclusive QCD.
Jets are reconstructed using Delphes EFlow module, with anti-$k_\mathrm{T}$ radius parameter $R=0.8$, and are required to have $p_\mathrm{T} \in [550, 650]$ GeV and $|\eta|<2$.
For each jet, we record the 4-momenta for the leading 140 particle-flow constituents.

\subsection{Architectures}
In previous sections, the discussion has focused on projection and convolution of a jet with a single filter $\Phi$.
In practice, we allow a neural network to learn a moderate number of independent filter channels $\Phi_c(r, \theta) = \sum_m \rho_{cm}(r) e^{i m \theta}$ and feature embeddings $Z_q$.
By convolving each feature embedding with each filter, the network produces a convolution waveform $\langle \Phi_c | R(\Delta) | Z_j\rangle$ with $C \times J$ feature channels, which can be passed on as multi-feature input to a standard 1D CNN architecture.
Hence, to define the architecture of a particular rPCN, we must to specify the following:
\begin{itemize}
    \item The number of filter channels, $C$;
    \item The number of feature embeddings, $J$;
    \item The maximum angular frequency mode, $M$;
    \item The feature embedding function(s), $Z_j(q)$;
    \item The radial activation function(s), $\rho_{cm}(r)$;
    \item The remaining 1D convolutional network structure.
\end{itemize}

Except in the IRC-safe case, we implement both $Z_j$ and $\rho_{cm}$ as densely-connected neural networks.
Therefore, $Z_j$ and $\rho_{cm}$ are specified by the number of hidden layers and their units, as well as their nonlinearies.
We apply a ReLU nonlinearity at the output of $Z_c$, so that it can more naturally act as an ``attention'' mechanism which can learn to ignore particles based on their scalar features $q_k$ in the region where the ReLU response is zero.
The complex-valued network $\rho_{cm}$ is implemented by a single network which learns real and imaginary parts separately, $\rho_{cm} = \alpha_{cm} + i \beta_{cm}$.
We impose the constraint $\rho_{-m} = \rho_m^*$ so that $\Phi_c$ are real by construction; therefore it is sufficient to learn $\alpha_{cm}$ for $m\geq 0$ and $\beta_{cm}$ for $m\geq 1$.

When particle ID is included as part of the scalar input, we follow the ``experimentally plausible'' labeling scheme of \cite{Komiske:2018cqr}, where particles are categorized into one of eight types: photons, neutral hadrons, and positively- and negatively-charged muons, electrons, and hadrons.
These labels are input to the network via a trainable 3-dimensional embedding layer, which are concatenated with the particle's $p_\mathrm{T}$ and passed as a triplet into the $Z_c$ network.

In the IRC-safe case, the function $Z_c$ is simply replaced with the transverse momentum of each particle.
Since the majority of trainable parameters in the Particle Convolution network tend to be from the $Z_c$ and $\rho_{cm}$ sub-networks, the IRC-safe networks will usually have substantially fewer parameters than their non-IRC-safe counterparts.

In our experiments, the remainder of the network can be specified by:
\begin{itemize}
    \item The nonlinearity following the Particle Convolution layer;
    \item The number of 1D convolution layers, their kernel size, and stride;
    \item The global pooling operation (maximum or average);
    \item Any remaining hidden layers, units, and nonlinearity;
    \item A final 2-unit softmax output layer representing the categorical prediction.
\end{itemize}

We found that a resnet-like~\cite{2015arXiv151203385H} convolutional architecture worked best.
We pass the particle convolution output without nonlinearity as a ``skip connection'' across residual blocks consisting of ReLU activation and batch normalization preceding two 1D CNN layers.

\subsection{Training}
In all experiments, the data are split into train, validation, and test samples in a 6:1:1 ratio.
Networks are implemented using Tensorflow v2~\cite{tensorflow2015-whitepaper} and Keras~\cite{chollet2015keras}.
The loss function used for training is the categorical crossentropy, optimized via Adam\cite{2014arXiv1412.6980K} with learning rate $10^{-4}$ and batch size $128$.
The validation loss is used to determine training convergence; training is stopped when the validation loss has not improved for $16$ consecutive epochs.
The model epoch with the lowest validation loss is evaluated on the test set to report unbiased AUC and rejection metrics.

After training networks by randomly sampling a wide range of hyperparameter configurations, we selected the best-performing architecture for each task based on the validation loss.
These configurations were then frozen, and the networks were retrained ten times.
The values reported in tables~\ref{tab:comparison-qg} and~\ref{tab:comparison-top} are the test scores for the specific network whose validation score was nearest to the median value on each task.
The errors quoted are the standard deviation of the test scores over the series of ten retrainings.

\subsection{Inspection}
\begin{figure}
    \centering
    \includegraphics[width=0.5\textwidth]{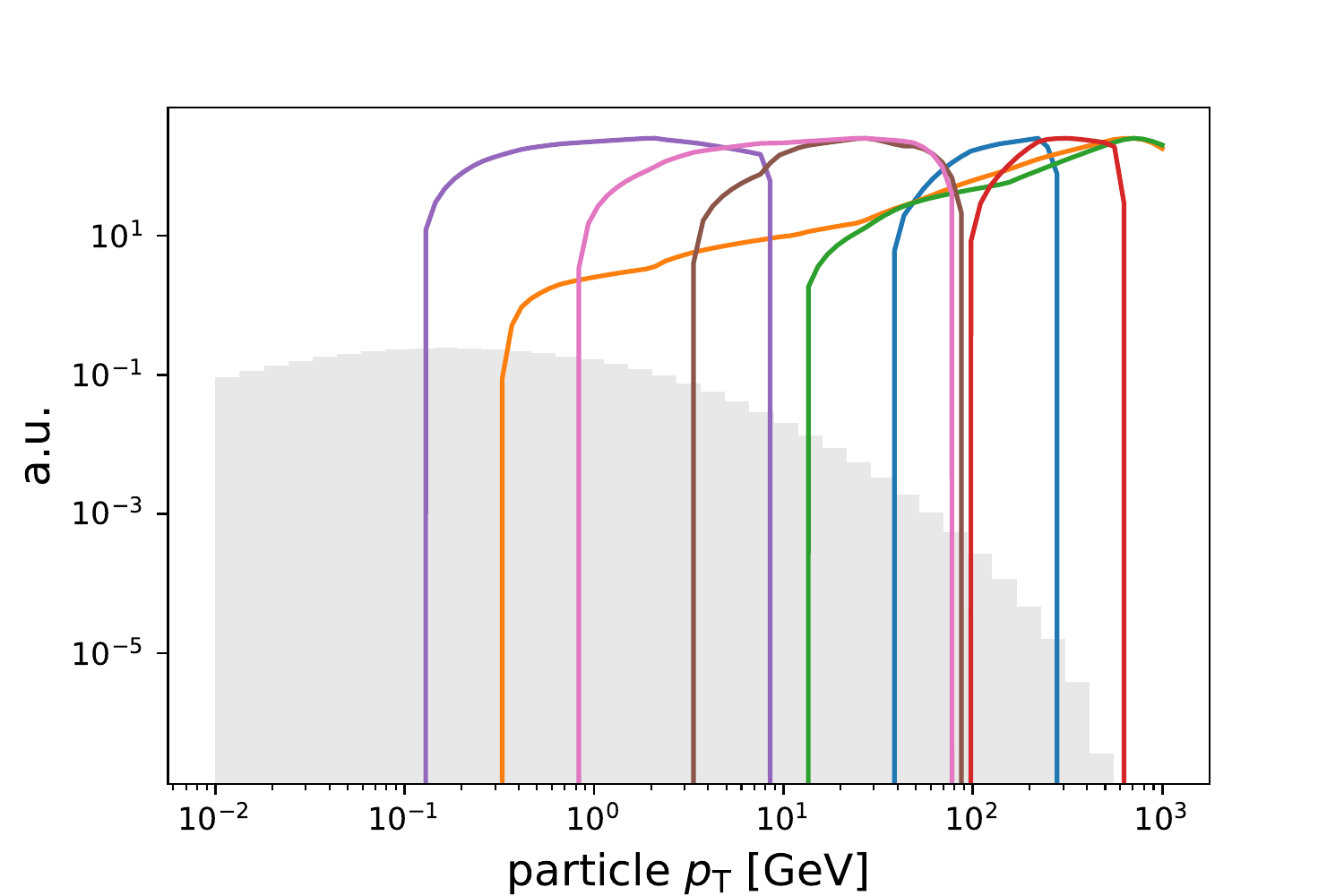}
    \caption{Colored lines representing functions $Z(p_\mathrm{T})$ learned by a network on the q/g tagging task. The shaded histogram indicates the distribution of particle $p_\mathrm{T}$ over the training dataset.}
    \label{fig:z-response}
\end{figure}

It is possible to visualize the learned functions $\Phi_c(\vec{x})$ in the $\eta - \phi$ plane.
Moreover, we can visualize the learned behavior of the nonlinear feature embedding, $Z_j(p_\mathrm{T})$.
Each convolution represents a learned spatial pattern as well as a gated response based on momentum.
Interestingly, it seems that the network often spontaneously learns a sort of binning in $p_\mathrm{T}$, as shown in Figure~\ref{fig:z-response}.
It also tends to ignore many of the lowest-$p_\mathrm{T}$ particles, suggesting those particles could potentially be pruned from the input to further speed up the network.

\section{Results \& Conclusions}
\label{sec:conclusion}

The rPCN performance on the two benchmark tasks are given in Tables~\ref{tab:comparison-qg} and \ref{tab:comparison-top}.
These results show that by promoting the permutation-invariant architectures EnergyFlow and ParticleFlow to include rotational convolution significantly improves their performance.
In particular, the rPCN significantly improves the state-of-the-art for IRC-safe taggers.
When IRC safety can be dispensed with, the rPCN can achieve performance comparable to the current state-of-the-art set by graph-based models such as ParticleNet.

From a practical perspective, the rPCN is much more similar to deep-sets based architectures already in production use by LHC experiments, as compared to the relatively novel graph architectures.
The rPCN model also requires minimal preprocessing, is compatible with arbitrary-length collections of particles, and is permutation-equivariant.
PCNs in general are conceptually analogous to the jet image approach in the limit of infinitely small pixels and continuous filter kenels, although the rPCN targets rotational rather than translational equivariance as a physically-motivated inductive bias.

Moreover, by careful design of the filter representation $\Phi$, the PCN approach can admit further generalization by constructing convolutional operations that can endow additional equivariance properties to jet tagging models.
For example, in forthcoming work we study whether a PCN which is equivariant \textit{w.r.t.} to scaling in the $\eta - \phi$ plane could potentially be useful to capture similar substructure properties across a wide range of jet $p_\mathrm{T}$.

\section{Acknowledgements}
The author would like to especially thank Paul Tipton for many iterations of feedback throughout this project.
He thanks Aishik Ghosh, Daniel Whiteson, Dan Guest, and Ema Smith for helpful conversations and comments on this manuscript.
Training experiments were made possible by the Grace GPU cluster operated by Yale Center for Research Computing. 
This work was supported by grant DE-SC0017660 funded by the U.S. Department of Energy, Office of Science.

\begin{table*}[ht]
    \centering
    \begin{tabular}{l r c c}
    \vspace{0.25 em}
    & Model & AUC & $1/\epsilon_b |_{\epsilon_s = 50\%}$ \\
    \hline
    \hline
    \multicolumn{1}{l}{\textit{IRC-safe}} & & & \\
    & EFN & 0.8824 & $28.6 \pm 0.3$ \\
    & rPCN \textbf{(Ours)} & \textbf{0.8944} & $\mathbf{32.5 \pm 0.4}$ \\
    \hline
    \multicolumn{1}{l}{\textit{w/o PID}} & & & \\
    & PFN & 0.8911 & 30.8 $\pm 0.4$ \\
    & ResNeXt-50 & 0.8960 & 30.9 \\
    & P-CNN & 0.8915 & 31.0 \\
    & ParticleNet-Lite & 0.8993 & 32.8 \\
    & ParticleNet & \textbf{0.9014} & 33.7 \\
    & rPCN \textbf{(Ours)} & 0.8997 & $\mathbf{34.2 \pm 0.4}$ \\
    \hline
    \multicolumn{1}{l}{\textit{w/ PID}} & & & \\
    & P-CNN & 0.9002 & 34.7 \\
    & PFN-Ex & 0.9005 & 34.7 \\
    & ParticleNet-Lite & 0.9079 & 37.1 \\
    & rPCN +Ex \textbf{(Ours)} & 0.9081 & $38.6 \pm 0.5$ \\
    & ParticleNet & \textbf{0.9116} & $\mathbf{39.8 \pm 0.2}$ \\
    \hline
    \end{tabular}
    \caption{
    Comparison of known network architectures on the quark-gluon tagging dataset of Ref.~\cite{Komiske:2018cqr}.
    ResNeXt is a deep 2D CNN adapted for jet images in~\cite{Qu:2019gqs}, while P-CNN~\cite{CMS-DP-2017-049} is a 1D convolution operating on ordered lists of particles, also implemented in~\cite{Qu:2019gqs}.
    Our IRC-safe rPCN model is substantially more sensitive than the EFN.
    When including a nonlinear $p_\mathrm{T}$ function, the rPCN model achieves comparable performance to ParticleNet.
    }
    \label{tab:comparison-qg}
\end{table*}

\begin{table*}[ht]
    \centering
    \begin{tabular}{l  r c c c}
    \vspace{0.25 em}
    & Model & AUC & $1/\epsilon_b |_{\epsilon_s = 50\%}$ & $1/\epsilon_b|_{\epsilon_s = 30\%}$ \\
    \hline
    \hline
    \multicolumn{1}{l}{\textit{IRC-safe}} & & & & \\
    & EFPs & 0.9803 & 184 & 384 \\
    & EFN & 0.9789 & 181 & 619 \\
    & rPCN \textbf{(Ours)} &  $\mathbf{0.9821}$ & $\mathbf{257 \pm 6}$ & $\mathbf{1038 \pm 41}$ \\
    \hline
    \multicolumn{1}{l}{\textit{w/o PID}} & & & & \\
    & P-CNN & 0.9803 & 201 & 759 \\
    & PFN & 0.9819 & 247 & 888 \\
    & ResNeXt-50 & 0.9837 & 302 & 1147 \\
    & ParticleNet-Lite & 0.9844 & 325 $\pm 5$ & 1262 $\pm 49$ \\
    & rPCN \textbf{(Ours)} & 0.9845 & 364 $\pm 9$ & $\mathbf{1642 \pm 93}$ \\ 
    & ParticleNet & $\mathbf{0.9858}$ & $\mathbf{397 \pm 7}$ & $1615 \pm 93$ \\
    \hline
    \end{tabular}
    \caption{
    Comparison of similar network architectures on the top tagging dataset of Ref.~\cite{Komiske:2018cqr}.
    ResNeXt is a deep 2D CNN adapted for jet images in~\cite{Qu:2019gqs}, while P-CNN~\cite{CMS-DP-2017-049} is a 1D convolution operating on ordered lists of particles, also implemented in~\cite{Qu:2019gqs}.
    As with the q/g task, the rPCN sets a new state-of-the-art for IRC-safe tagging, while achieving comparable performance to the graph-based ParticleNet model.
    }
    \label{tab:comparison-top}
\end{table*}

\newpage

\bibliographystyle{unsrt}  
\bibliography{references}  

\end{document}